\documentstyle[12pt,twoside,fleqn,espcrc1,epsfig]{article}
\newcommand{\AmS}{{\protect\the\textfont2
  A\kern-.1667em\lower.5ex\hbox{M}\kern-.125emS}}

\begin{document}
% declarations for front matter
\title{Particle   Ratios   at  CERN,  BNL  and  GSI  :  Unified
description of Freeze-Out Parameters.\footnote{Presented at Quark Matter 99, Torino, Italy}}

\author{J. Cleymans\address{Department of Physics, 
        University of Cape Town, \\ 
        Rondebosch 7701, South Africa}%
        \thanks{On sabbatical leave at the Fakult\"at f\"ur Physik,
	Universit\"at Bielefeld, D-33615 Bielefeld, Germany.}
        and 
        K. Redlich\address{Department of Theoretical Physics, 
	University of Wroclaw, \\
        PL-50204 Wroclaw, Poland}
        \thanks{On leave of absence at the GSI,
	D-64291 Darmstadt, Germany.}}

%\begin{document}
% typeset front matter
\maketitle
\begin{abstract}
It is shown that the chemical freeze-out parameters obtained at CERN/SPS,
BNL/AGS and GSI/SIS energies all correspond to a unique value of 1 GeV for the 
average energy per hadron. 
\end{abstract}
\section{Particle Ratios : General Remarks}
Information about the chemical freeze-out
parameters,  
the temperature, $T_{ch}$, and  
baryon chemical potential $\mu_B^{ch}$ can be obtained from
ratios of integrated particle yields \cite{heinz,padova}. This
is because various effects like transverse flow
or particle production from a superposition of fireballs cancel out
in such   ratios
provided the freeze-out parameters are unique. We will discuss 
in succession particle ratios
at GSI/SIS, BNL/AGS and SPS/CERN. We will then discuss  common properties,
in particular, the observation 
that all of them correspond to an average energy of 1 GeV per hadron 
independent of the beam energy \cite{prl}.

\section{Particle ratios at GSI/SIS}

\begin{figure}[htb]
%\begin{minipage}[t]{180mm}
\begin{center}
\epsfig{file=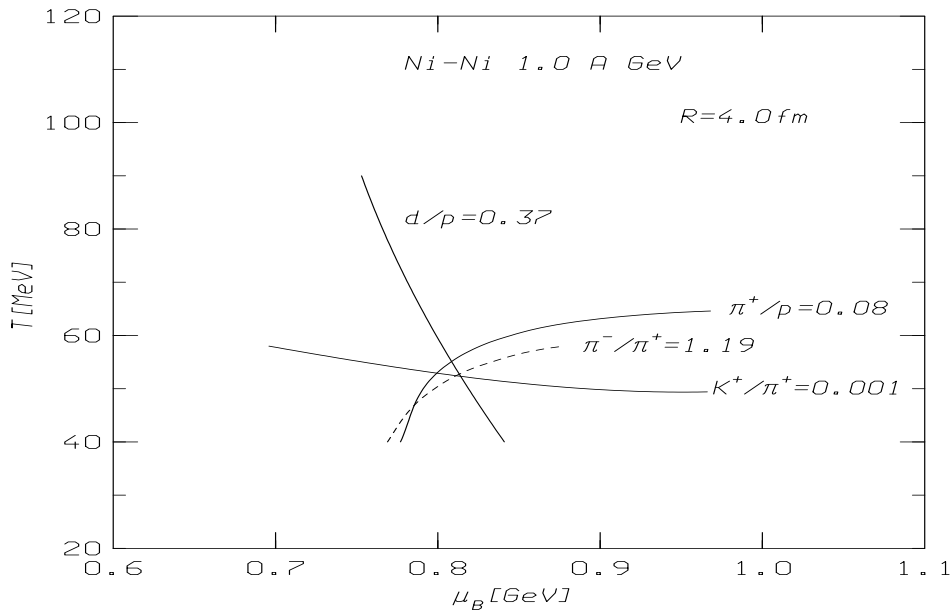, width=125mm,height=80mm}
\end{center}
\vspace*{-1.2cm}
\caption{Particle ratios for Ni-Ni at 1.0 A GeV at SIS
\cite{oeschler}.}
\label{fig:GSI}
%\end{minipage}
\vspace*{-0.5cm}
\end{figure}

A systematic study of the  particle ratios
measured  at GSI/SIS
energies has been done recently \cite{keranen,oeschler}.  
In 
figure \ref{fig:GSI} we show 
the results obtained in \cite{oeschler} 
for Ni-Ni at 1.0 AGeV. 
The particle ratios  are
consistent with the interpretation that the hadronic composition
of the final state 
is fixed
at a unique temperature and baryon chemical potential.
Since the temperature is low, the number of particles created is
small and it is therefore necessary to take into account
the exact conservation of strangeness because strange particles
are always produced in pairs and it is more difficult to create two
particles in a small cold system than it is to create one particle.
There is no necessity to introduce any other  parameters 
and  a good fit can be achieved with full chemical equilibrium
including strange particles, i.e.
with $\gamma_s = 1$. In view of the fact that kaons are produced below
threshold at SIS this is remarkable.
\section{Particle ratios at BNL/AGS}
\begin{figure}[htb]
%%\begin{minipage}[t]{80mm}
\begin{center}
\epsfig{file=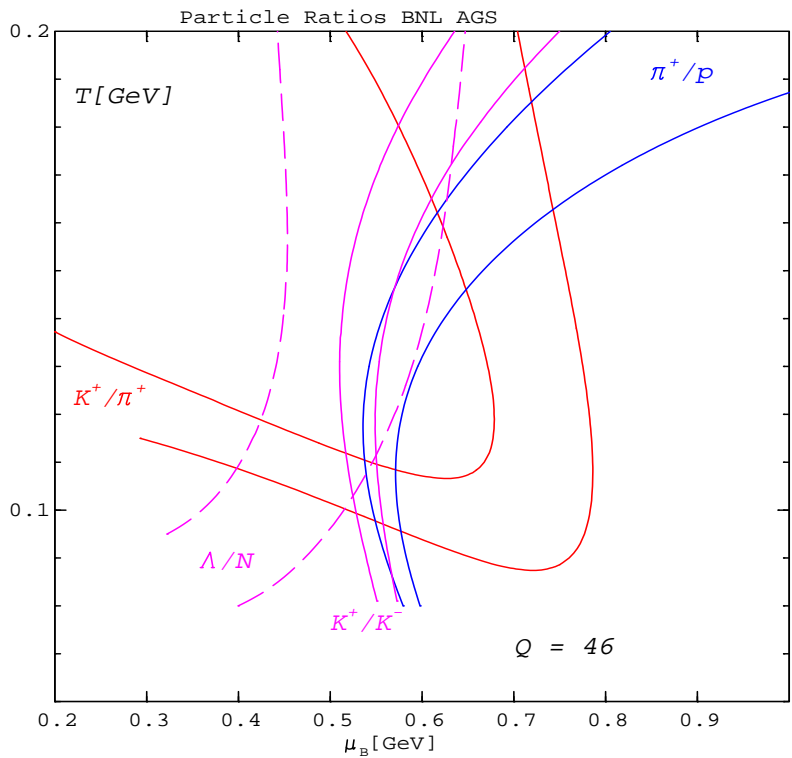, width=150mm,height=120mm}
\end{center}
\vspace*{-1.4 cm}
\caption{Particle ratios for Si-Au at 11.0 A GeV at AGS \cite{elliott}.}
\label{fig:AGS}
%\end{minipage}
\end{figure}

Particle ratios at BNL/AGS have been reanalyzed recently 
in references \cite{becattini} and \cite{heppe2}. The results
are consistent with those obtained  previously, 
e.g. ref. \cite{becattini} obtains $T_{ch}=118.4\pm11.6$ MeV
and $\mu_B^{ch} \approx 522$ MeV. 
The ratios together with error bands are shown in figure \ref{fig:AGS}
\cite{elliott}.
This temperature is  about double the one extracted from SIS results,
the baryon chemical potential is clearly much smaller. 
Because of the higher temperature,
it is no longer necessary to treat strangeness in a special way
and  corrections from the more exact canonical treatment are negligible.
 There is agreement 
that a good fit can be achieved
with $\gamma_s = 1$
and  the data are again consistent with chemical equilibrium.
\vspace*{-0.5cm}
\section{Particle ratios at CERN/SPS}
Several papers have appeared recently analyzing the particle ratios
measured in Pb-Pb collisions at CERN \cite{heppe,gorenstein,rafelski}.
In reference \cite{gorenstein} 
it is found that
a strangeness
suppression factor $\gamma_s = 0.55$  is needed  to 
describe the data while in reference \cite{heppe}  
it is argued that a good description is possible using
$\gamma_s = 1$.
Despite this serious difference both papers arrive at very similar
results : $T_{ch}=165$ MeV in \cite{gorenstein} vs $T_{ch}=170$ MeV in
\cite{heppe}, i.e. the difference is less than 3 percent.
The disagreement over the value of $\gamma_s$ will be settled when more
precise data become available, indications from this
 conference are that a value below
1 is necessary but a full analysis is still outstanding.

The analysis of reference \cite{rafelski} introduces  new
parameters and is  
not directly comparable to the one discussed here.
\section{Particle Ratios : Common Properties at SPS, AGS and SIS}
The results from GSI, BNL and CERN 
show a striking systematic behavior (see figure 3) : the GSI/SIS 
results have the lowest
temperature and the highest baryon chemical potential, as the beam energy
is increased a clear shift towards higher temperatures and
lower baryon chemical potentials occurs. The
points have in common that the average 
energy per hadron is approximately 1 GeV \cite{prl}. 
Chemical freeze-out
is thus reached when  
the energy per particle drops
below 1 GeV per hadron.
When this value is reached 
inelastic collisions are no longer important 
and the abundances of the various hadronic species
are fixed. 
The consequences of this are discussed in detail in reference
\cite{preprint}. Recent results from E895 collaboration presented at this conference \cite{rai}
are compatible with the above result.

\begin{figure}[htb]
%\begin{minipage}[b]{120mm}
\begin{center}
\epsfig{file=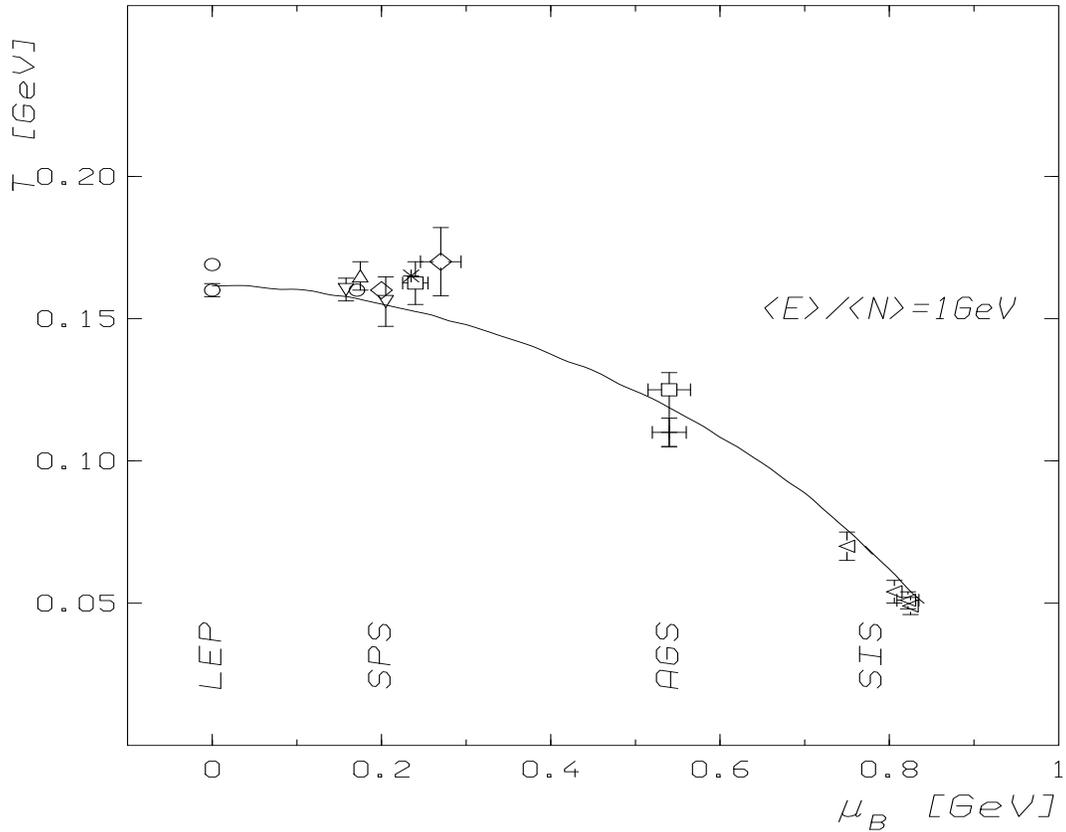, width=140mm,height=110mm}
\end{center}
\vspace*{-1.2cm}
\caption{Freeze-out values obtained from ratios of hadronic abundances.
The smooth curve corresponds to a fixed energy per hadron 
of 1 GeV in the hadronic
gas model.}
\label{fig:freeze}
%\end{minipage}
\end{figure}
\vspace*{-0.5cm}
\section{Conclusions}
Bringing together results obtained at very different beam energies shows that 
the hadronic abundances seen in the final state of relativistic heavy ion 
collisions are fixed once the average energy per hadron drops below 1 GeV.
We expect the results from the SPS beams at 40 and 80 GeV to follow this 
observation.
It will be interesting to see how the results from RHIC will relate to this
 observation.

We thank P. Braun-Munzinger, B. Friman, W. N\"orenberg, H. Oeschler,
H. Satz and J. Stachel for fruitful discussions.

\end{document}